\documentclass[a4paper,11pt]{article}
\usepackage{pos}
\usepackage{adjustbox}
\usepackage{multirow}
\usepackage{graphicx}
\usepackage{natbib}
\title{TeV and optical observations of the Be/pulsar binary 1A 0535+262 during the 2020 giant outburst}
\ShortTitle{VERITAS 1A 0535+262}

\author*[a]{Matthew Lundy}
\forColl{VERITAS} 

\affiliation[a]{Department of Physics, McGill University,\\
  Montreal, QC H3A 2T8, Canada}

\emailAdd{matthew.lundy@mail.mcgill.ca}

\abstract{

1A 0535+262 is a Be X-ray binary pulsar and one of the only galactic pulsar systems to show radio jet emission. Characterizing the very high energy emission (VHE, >100 GeV) in these extreme microquasars is critical to understanding their contribution to the origin of galactic cosmic rays. The 2020 giant outburst of this system, where X-ray fluxes exceeded 12 Crab, marked a rare opportunity to investigate the gamma-ray and rapid optical variability of these transient systems while in such an extreme state. This month of activity marked the brightest flare measured in this system. VERITAS's developing optical capabilities in tandem with the ability to measure TeV gamma rays allowed for a unique campaign to be undertaken. VERITAS's observations of this system during the outburst will be presented in the context of observations at lower energies and previous observations of this system by imaging atmospheric Cherenkov telescopes.}

\FullConference{37$^{\rm{th}}$ International Cosmic Ray Conference (ICRC 2021)\\
		July 12th -- 23rd, 2021\\
		Online -- Berlin, Germany}

\begin{document}
\maketitle

\section{Introduction}
Galactic binaries have proven one of the most interesting source classes for the current generation of ground based gamma-ray telescopes. Transient flares in these systems, like those found in LS I +61 303, have been of particular interest and a general overview of VERITAS's observations of these systems can be found in an accompanying ICRC proceeding. 1A 0535+262 is a high mass X-ray binary system (HMXB) with an orbital period of $\sim$111 days \cite{2006MNRAS.368..447C} and a pulsar rotational period of $\sim103$ s \cite{1975Natur.256..628R}. The system is composed of a O9.7-B0 IIIe star and a neutron star located at a distance of $\sim 2$ kpc \cite{1998MNRAS.297L...5S}.  Type II giant outbursts from 1A 0535+262 are seen roughly every 5 years. In previous flares, Swift-BAT observed fluxes over 1 count/cm$^2$/s exceeding four times the flux of the Crab (in the 15-50 keV band). The flare occurring in November 2020 exceeded an intensity level of 10.5 Crab in Swift-BAT making it one of the most extreme flares seen in a nearby HMXB. Prior to this outburst the brightest 1A 0535+262 flare was observed by BATSE (20-40keV) with a peak flux of over 8 Crab in 1994 \cite{2012ApJ...754...20C}. This latest flaring event triggered follow-up from a large number of X-ray telescopes including Nicer/NuSTAR \cite{2020ATel14179....1J}, and continued monitoring in MAXI/GSC \cite{2020ATel14173....1N} and Swift/BAT \cite{2020ATel14170....1P}.

Radio follow-up with VLA at 6 GHz triggered by this event showed a brightening radio counterpart rising from 13 to 39 $\pm 4$ $\mu$Jy during the rising phase of the X-ray flare \cite{2020ATel14193....1V}. Radio emission was not seen in previous flares from 1A 0535+262. This radio emission is normally attributed to the formation of a relativistic jet which would be unexpected in the high-magnetic fields surrounding a neutron star. The only comparable system is the 2017 flare in the HMXB Swift J0243+624 observed in 2017 \cite{2018Natur.562..233V}. Measured X-ray fluxes in both of these systems also suggest that during these epochs super-critical accretion near or above the spherical Eddington limit occurred \cite{2021ATel14392....1M}. Investigating the potential gamma-ray emission during these stages can help to constrain this novel and extreme environment.

1A 0535+262 has been associated with the EGRET unidentified gamma-ray source 3EG J0542+2610 \cite{2001A&A...376..599R} however the current generation of instruments has been unable to replicate this result. VERITAS has previously observed and reported on a flare that occurred in 2011 \cite{Acciari_2011}. During this time VERITAS and \textit{Fermi}-LAT did not detect significant VHE or HE emission from the system.  

Accreting compact objects also provide exciting environments to probe with rapid optical photometry. Cadences on the millisecond timescale are necessary to fully probe the flux variation in many of these systems. Combining this data with rapid X-ray data and investigating correlations between X-ray and optical variations probes the lowest base of the jet and the interaction with accretion processes \cite{2013MNRAS.429L..20M}. Observations of this style have already been performed on black hole X-ray binaries during outbursts for example; MAXI-J1820+070 \cite{2019MNRAS.490L..62P},  GX 339-4 \cite{2008MNRAS.390L..29G} and V404 Cyg \cite{2016MNRAS.459..554G}. Observations have shown sub-second correlations between the X-ray and optical observations and a wavelength dependent lag that provides a measurement of the ms lag between the X-ray corona and the emission region of the jet. The sporadic nature of flares in these systems combined with the limited number of observatories capable of rapid optical photometry, has resulted in a limited sample of systems studied. 

In this proceeding we will present the results from the VERITAS optical and gamma-ray observations of 1A 0535+262 during the Type II outburst in November 2020.

\section{Instrumentation }
VERITAS is an array of four Imaging Atmospheric Cherenkov Telescopes (IACTs) located at the Fred Lawrence Whipple Observatory (FLWO) in southern Arizona (31 40N, 110 57W,  1.3km a.s.l.) \cite{2008AIPC.1085..657H}. Each telescope is comprised of 499 photomultiplier tubes (PMTs) covering a field of view of $\sim3.5^\circ$. VERITAS detects gamma-ray photons from 100 GeV to >30 TeV and can detect a source with 1\% Crab in 25 hours \cite{2015ICRC...34..771P}. In 2018, VERITAS's Enhanced Current Monitor (ECM) system was installed. The ECM is a parasitic backend that allows for optical observations to take place during regular observations using the VERITAS camera. Due to the large integrated NSB in the 0.15$^\circ$ pixel monitored by the ECM, the limiting magnitude of the VERITAS instrument is $\sim12$ mag. This means that only bright sources can be observed by the ECM. For optical observations we do not filter the light entering the PMT, meaning that the band monitored is simply a factor of the wavelength-dependent quantum efficiency of the PMT. For more details on the configuration see the VERITAS FRB ICRC proceeding from this conference.  
\section{Observations}

\begin{figure}
    \centering
    \includegraphics[width=14cm]{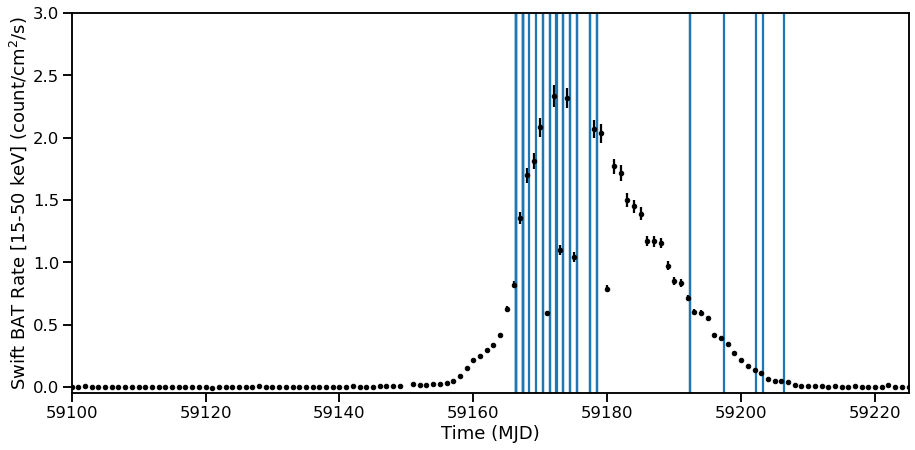}
    \caption{Times of VERITAS Observations (in blue) compared to the Swift-BAT X-ray flux.}
    \label{fig:obs}
\end{figure}

The VERITAS observations of 1A 0535+262 during the 2020 season totals 15.5 hours of gamma-ray and optical data. The triggering criteria, based on the estimated optical brightness of the source, for VERITAS's program was met on 2020-11-13 and continued until 2020-12-23. Observations were taken during dark sky and low moonlight condition between 50-85 degree elevation. Data was broken up into 30 minute runs taken in "ON" mode, with the source centered in the camera allowing for ECM observations. Initial observations were taken daily but as the flare evolved the cadence was reduced and days overlapping NICER windows were prioritized. A comparison of the times of VERITAS observations and the X-ray flux can be seen in Figure \ref{fig:obs}.  

\section{Analysis and Results}
\begin{figure}
    \centering
    \includegraphics[width=14cm]{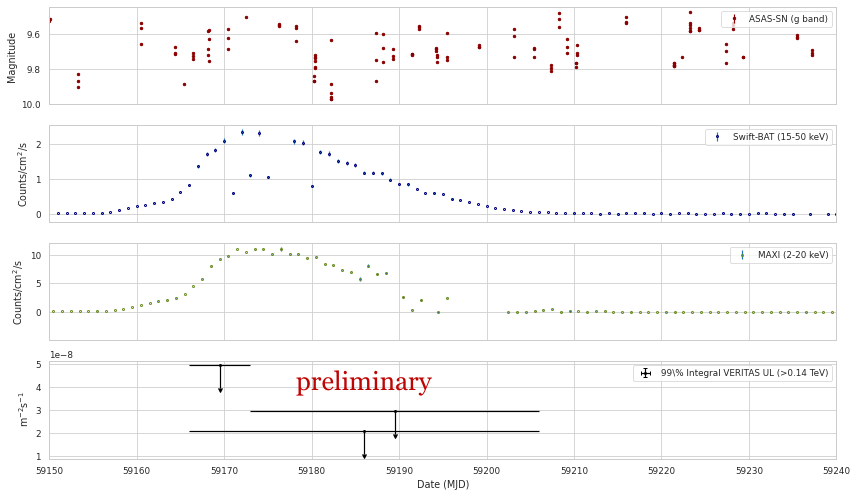}
    \caption{Optical, X-ray and VHE lightcurves during the 2020 Type II outburst of 1A 0535+262.}
    \label{fig:lc}
\end{figure}

The gamma-ray data was analyzed using the standard VERITAS software package \cite{2008ICRC....3.1385C}, with cuts optimized for soft sources (spectral index of -3.0 $\longrightarrow$ -4.0). The significance was calculated with off-source rates found using the ring-background method. The system is expected to be opaque to gamma-rays near the peak of the outbursts due to thermal emission from the accretion disk providing a large population of ambient photons. An analysis of the entire time range was performed as well as individual investigations of the rising and falling slopes. There was no significant detection of gamma-rays in any of these time windows. Upper limits on the VHE flux compared to the X-ray flux can be seen in Figure \ref{fig:lc}.

\begin{figure}
    \centering
    \includegraphics[width=15cm]{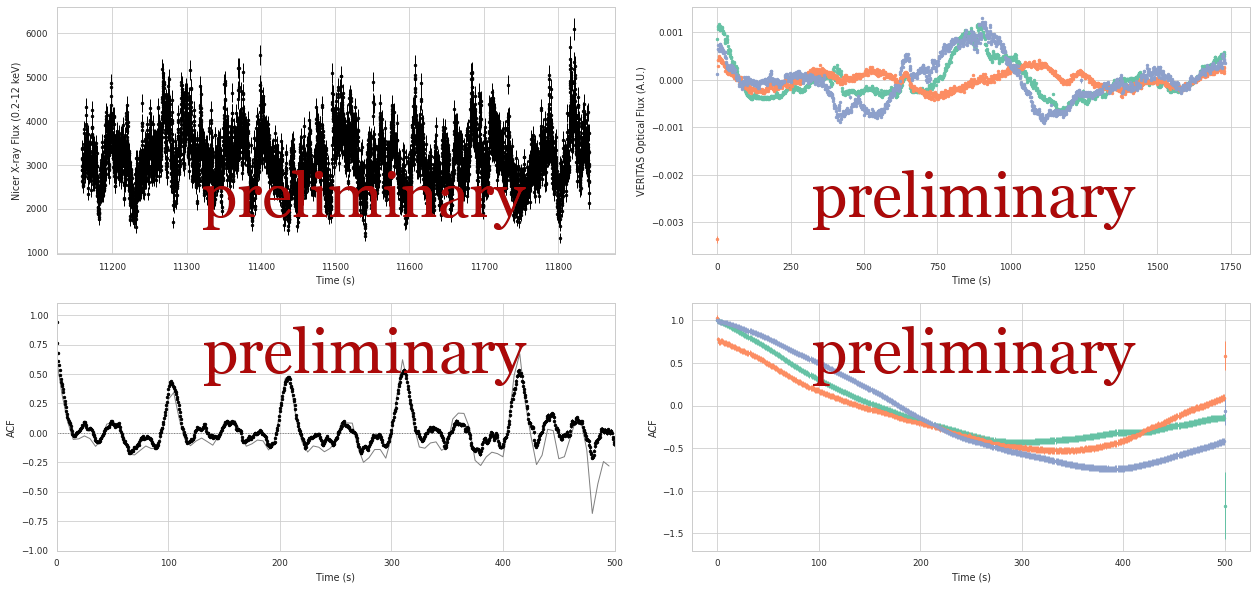}
    \caption{NICER X-ray (top-left), and VERITAS ECM optical (top-right) light curves for 1A 0535+262 taken on 2020-11-14. Three VERITAS telescopes are plotted with distinct colors to separate them (blue:2, orange:3, green:4). Below the light curves are the respective auto-correlation functions for the above light curve. Autocorrelations were calculated for the NICER data with both the Scargle (line) and Edelson and Krolik method (points). For the VERITAS data only the Edelson and Krolik method is shown. }
    \label{fig:autocorr}
\end{figure}

The optical data was investigated for periodic signals seen in the simultaneous X-ray data from the NICER telescope (the 103 s period was of particular interest). NICER data was taken from 0.2-12 keV, and good time intervals were taken from the preprocessed data available on the HEASARC archive \cite{2012AAS...21924905A}. Discrete correlation functions between the X-ray and optical were calculated as well as autocorrelations for all runs with contemporaneous NICER data. A preliminary investigation of Lomb-Scargle and Fourier periodograms showed no strong periodic signals in the VERITAS optical data near signals identified in the X-ray. An investigation of non-periodic small timescale correlations (i.e. flares) is underway along with other additional investigations of potential correlated behavior. The auto-correlation function for the data taken on 2020-11-14 is shown in Figure \ref{fig:autocorr}.

\section{Conclusion and Outlook}

This campaign is one of the first attempts at joint gamma-ray and optical observations of a gamma-ray binary using a IACT. HMXBs are one source class where having both the optical and gamma-ray data can prove useful. VERITAS continues to search for other potential sources that leverage both sides of the IACT instrument including M-dwarfs and FRBs. Further investigation of the optical data presented here will be included in a future publication.
\section{Acknowledgments} 

\noindent For VERITAS see acknowledgments see: \url{https://veritas.sao.arizona.edu/}

\noindent Additionally this research was undertaken thanks in part to funding from the Canada First Research Excellence Fund through the Arthur B. McDonald Canadian Astroparticle Physics Research Institute.

\medskip

\small
\bibliography{bib}

%% Full authors list (ONLY FOR COLLABORATIONS)
\clearpage
\section*{Full Authors List: \Coll\ Collaboration}

\scriptsize
\noindent
C.~B.~Adams$^{1}$,
A.~Archer$^{2}$,
W.~Benbow$^{3}$,
A.~Brill$^{1}$,
J.~H.~Buckley$^{4}$,
M.~Capasso$^{5}$,
J.~L.~Christiansen$^{6}$,
A.~J.~Chromey$^{7}$, 
M.~Errando$^{4}$,
A.~Falcone$^{8}$,
K.~A.~Farrell$^{9}$,
Q.~Feng$^{5}$,
G.~M.~Foote$^{10}$,
L.~Fortson$^{11}$,
A.~Furniss$^{12}$,
A.~Gent$^{13}$,
G.~H.~Gillanders$^{14}$,
C.~Giuri$^{15}$,
O.~Gueta$^{15}$,
D.~Hanna$^{16}$,
O.~Hervet$^{17}$,
J.~Holder$^{10}$,
B.~Hona$^{18}$,
T.~B.~Humensky$^{1}$,
W.~Jin$^{19}$,
P.~Kaaret$^{20}$,
M.~Kertzman$^{2}$,
D.~Kieda$^{18}$,
T.~K.~Kleiner$^{15}$,
S.~Kumar$^{16}$,
M.~J.~Lang$^{14}$,
M.~Lundy$^{16}$,
G.~Maier$^{15}$,
C.~E~McGrath$^{9}$,
P.~Moriarty$^{14}$,
R.~Mukherjee$^{5}$,
D.~Nieto$^{21}$,
M.~Nievas-Rosillo$^{15}$,
S.~O'Brien$^{16}$,
R.~A.~Ong$^{22}$,
A.~N.~Otte$^{13}$,
S.~R. Patel$^{15}$,
S.~Patel$^{20}$,
K.~Pfrang$^{15}$,
M.~Pohl$^{23,15}$,
R.~R.~Prado$^{15}$,
E.~Pueschel$^{15}$,
J.~Quinn$^{9}$,
K.~Ragan$^{16}$,
P.~T.~Reynolds$^{24}$,
D.~Ribeiro$^{1}$,
E.~Roache$^{3}$,
J.~L.~Ryan$^{22}$,
I.~Sadeh$^{15}$,
M.~Santander$^{19}$,
G.~H.~Sembroski$^{25}$,
R.~Shang$^{22}$,
D.~Tak$^{15}$,
V.~V.~Vassiliev$^{22}$,
A.~Weinstein$^{7}$,
D.~A.~Williams$^{17}$,
and 
T.~J.~Williamson$^{10}$\\
\noindent
$^{1}${Physics Department, Columbia University, New York, NY 10027, USA}
$^{2}${Department of Physics and Astronomy, DePauw University, Greencastle, IN 46135-0037, USA}
$^{3}${Center for Astrophysics $|$ Harvard \& Smithsonian, Cambridge, MA 02138, USA}
$^{4}${Department of Physics, Washington University, St. Louis, MO 63130, USA}
$^{5}${Department of Physics and Astronomy, Barnard College, Columbia University, NY 10027, USA}
$^{6}${Physics Department, California Polytechnic State University, San Luis Obispo, CA 94307, USA} 
$^{7}${Department of Physics and Astronomy, Iowa State University, Ames, IA 50011, USA}
$^{8}${Department of Astronomy and Astrophysics, 525 Davey Lab, Pennsylvania State University, University Park, PA 16802, USA}
$^{9}${School of Physics, University College Dublin, Belfield, Dublin 4, Ireland}
$^{10}${Department of Physics and Astronomy and the Bartol Research Institute, University of Delaware, Newark, DE 19716, USA}
$^{11}${School of Physics and Astronomy, University of Minnesota, Minneapolis, MN 55455, USA}
$^{12}${Department of Physics, California State University - East Bay, Hayward, CA 94542, USA}
$^{13}${School of Physics and Center for Relativistic Astrophysics, Georgia Institute of Technology, 837 State Street NW, Atlanta, GA 30332-0430}
$^{14}${School of Physics, National University of Ireland Galway, University Road, Galway, Ireland}
$^{15}${DESY, Platanenallee 6, 15738 Zeuthen, Germany}
$^{16}${Physics Department, McGill University, Montreal, QC H3A 2T8, Canada}
$^{17}${Santa Cruz Institute for Particle Physics and Department of Physics, University of California, Santa Cruz, CA 95064, USA}
$^{18}${Department of Physics and Astronomy, University of Utah, Salt Lake City, UT 84112, USA}
$^{19}${Department of Physics and Astronomy, University of Alabama, Tuscaloosa, AL 35487, USA}
$^{20}${Department of Physics and Astronomy, University of Iowa, Van Allen Hall, Iowa City, IA 52242, USA}
$^{21}${Institute of Particle and Cosmos Physics, Universidad Complutense de Madrid, 28040 Madrid, Spain}
$^{22}${Department of Physics and Astronomy, University of California, Los Angeles, CA 90095, USA}
$^{23}${Institute of Physics and Astronomy, University of Potsdam, 14476 Potsdam-Golm, Germany}
$^{24}${Department of Physical Sciences, Munster Technological University, Bishopstown, Cork, T12 P928, Ireland}
$^{25}${Department of Physics and Astronomy, Purdue University, West Lafayette, IN 47907, USA}

\end{document}